\documentclass[a4paper,12pt]{article}
 \usepackage[latin1]{inputenc}
\usepackage[T1]{fontenc}
\usepackage{amsmath}
\usepackage{amsfonts}
\usepackage{amssymb}
\usepackage{color}
\usepackage{graphicx}
\usepackage{caption}
\usepackage{subcaption}%

 \newcommand{\be}{\begin{equation}}
	 \newcommand{\ee}{\end{equation}}
	 \newcommand{\ba}{\begin{eqnarray}}
		 \newcommand{\ea}{\end{eqnarray}}
		 
		   \newcommand{\bea}{\begin{eqnarray}}
			 \newcommand{\eea}{\end{eqnarray}}
 \newcommand{\nn}{\nonumber}

\begin{document} 
\title{A Simple Model for a Dual Non-Abelian\\ Monopole -Vortex Complex}
\author{Gianni Tallarita$^{a}$ and Adam Peterson$^{b}$ \\\\
\normalsize \it  $^{a}$ Departamento de Ciencias, Facultad de Artes Liberales, Universidad Adolfo Ibáñez,\\ \normalsize \it Santiago 7941169, Chile\\\\
\normalsize \it  $^{b}$ Department of Physics, University of Toronto,\\ \normalsize \it Toronto, ON M5S 1A7, Canada
}
\date{\hfill}

\maketitle
\begin{abstract}

We investigate the flux-tube joining two equal and opposite electric charges using the dual Ginzburg-Landau model of superconductivity. The model is supplemented with an additional scalar field carrying a non-Abelian global symmetry, broken in the vortex cores. The presence of orientational moduli makes the flux tube non-Abelian. We perform a detailed study of the low energy theory of this soliton. We also analyze the solution representing superconducting droplets in the presence of the monopole anti - monopole pair.

\end{abstract}
\section{Introduction}

\quad One of the major problems of modern day particle physics is that of confinement of quarks. After decades of research a detailed comprehension of what exactly holds two quarks together is still lacking. This paper will not solve the problem. There have been however several ideas regarding what may be responsible for this mechanism and the general consensus, supplemented by lattice data, suggests some soliton structure is forming between quark charges which acts so as to confine them. Therefore, a flux tube of unknown origin or characteristics (presumably formed by degrees of freedom carried by the non-Abelian gauge field of QCD) sits between two electric monopole charges. In this way a monopole-vortex complex is formed. This is completely analogous (or dual) to what happens in conventional type II superconductors: two magnetic monopoles placed on opposite sides of a superconductor are confined by an Abrikosov vortex which forms between them. This idea of confinement by formation of an electric flux soliton is known as the dual mechanism of colour confinement \cite{'tHooft:1981ht} \cite{Mandelstam:1974pi} \cite{Ripka:2003vv}. Even though the strong coupling of low energy QCD prevents us from knowing the detailed properties of this object (at least theoretically) there is one major thing that is known: the dual vortex cannot be of the Abrikosov type \cite{Yung:2000uy}. Also, since quarks are electric sources, the flux carried by the vortex should be electric and not magnetic. Therefore, we are led to find new kind of flux-tube solutions which might be of closer resemblance to real world confinement.  One such modification is the non-Abelian vortex \cite{Auzzi:2003fs} \cite{Hanany:2003hp}. This kind of flux-tube is similar to the Abrikosov one but carries additional degrees of freedom. They come from a pattern of symmetry breaking in the vortex core resulting from a scalar field, charged under a global non-Abelian group, condensing there. An extensive literature on this subject exists (see \cite{Shifman:2012zz} \cite{Shifman:2009zz} and references within) which mainly focuses on somewhat involved constructions, containing for the most part suspersymmetry. Only recently a simple model was put forward which reproduces these kinds of solitons in a transparent way \cite{Shifman:2012vv}.  It was used to extend many solitons to their non-Abelian counterpart \cite{Canfora:2016spb} \cite{Peterson:2015tpa} \cite{Shifman:2015ama} \cite{Peterson:2014nma} \cite{Shifman:2014oqa} and even in a holographic setup \cite{Tallarita:2015mca}. From a different perspective it also appears to play an important cosmological role, as observed in \cite{Forgacs:2016dby}. In this paper we will use it in the context of the dual model of superconductivity and the non-Abelian monopole - vortex complex. Previous studies, based on the aforementioned more involved models, of such soliton complexes exist \cite{Auzzi:2003em} \cite{Cipriani:2011xp}. In particular they involve the presence of t'Hooft - Polyakov monopole sources, and commonly magnetic flux. In this paper we will not take this approach and rather consider the non-Abelian flux tube confining two electric Dirac sources. It can be thought of as a non-Abelian generalization of \cite{Maedan:1989ju}. We will show that one can reproduce and study the characteristics of this non-Abelian soliton complex from this simpler perspective in a clear and elegant fashion. \\

The starting point is the action for the dual Ginzburg-Landau model of superconductivity (our considerations will be classical and assume all couplings are weak), 
\be\label{eq1}
S_A = \int d^4x \left(-\frac{1}{4}\bar{F}_{\mu\nu}\bar{F}^{\mu\nu}+\frac{1}{2}(D_\mu\psi)(D^\mu\psi)^*-\frac{1}{2}\lambda(|\psi|^2-v^2)^2\right),
\ee
where 
\be
\bar{F}_{\mu\nu} = \partial_\mu B_\nu - \partial_\nu B_\mu + \bar{G}_{\mu\nu},
\ee
is the dual field strength tensor described by the vector potential $B_\mu$. The term $\bar{G}_{\mu\nu}$ is the Dirac string term which accounts for the presence of electric charge sources. The covariant derivative is
\be
D_\mu \psi = \partial_\mu\psi + i g B_\mu \psi,
\ee
where $g$ can be viewed as a magnetic charge. The bar notation denotes duality in the usual electro-magnetic sense. Therefore the usual field strength tensor
\be
F_{\mu\nu} = \overline{\partial_\mu B_\nu - \partial_\nu B_\mu} + G_{\mu\nu},
\ee
where
\be
\partial_\nu G^{\mu\nu} = j^\mu,
\ee
with $j^\mu$ the electric current. The electric charge is $j^0 = \rho$, with the electric field $E_i = G_{0i}$ satisfying $\nabla \cdot \vec{E} = \rho$. \newline

The energy for static field configurations is (we set $B_0 = 0$) 

\be\label{energy}
E = \int d^3x \left[\frac{1}{2}\left(-\vec{\nabla}\times\vec{B}+\vec{E}_{st}\right)^2+\frac{1}{2}(D_i \psi)(D^i \psi)^*+\frac{\lambda}{2}(|\psi|^2-v^2)^2\right].
\ee
where we have introduced the dual string term $\bar{E}^{st}_i=\bar{G}_{0i}$. \\

In this paper we are interested in the charge density caused by two equal and opposite electric charges, of charge $e$, separated in the $z$ direction.  We take the charges in the probe limit and assume they have infinite mass.  Therefore we will take
\be
\rho = e \left(\delta^3(\vec{r}-\vec{R}_1)-\delta^3(\vec{r}-\vec{R}_2)\right)
\ee
with 
\be
\vec{R}_1 = \left(0,0, -R/2\right), \quad \vec{R}_2 = \left(0,0, R/2\right).
\ee
This charge distribution produces the well known Coulomb field
\be
E^C_i = \frac{e}{4\pi} \left(\frac{(r-R_1)_i}{|r-R_1|^3}-\frac{(r-R_2)_i}{|r-R_2|^3}\right).
\ee

The Ball - Caticha procedure \cite{ballcatch} allows one to include the influence of the electric charges in the dual vortex energy minimizing equations. Following this we express the electric field in terms of the dual vector potential $B_0$ and the dual string term $\bar{E}^{st}_i=\bar{G}_{0i}$. The relation between the electric field and the dual variables is
\be
E^C_i = \bar{E}^{C_{st}}_i - (\nabla \times \vec{B}^C)_i.
\ee

The field $\vec{B}^C$ is inferred from the Dirac string expression (see Chapter 3 of \cite{Ripka:2003vv} for details and derivations) and can be written, in cylindrical coordinates, as
\be
\vec{B}^C(r) = \frac{e}{4\pi} \vec{\nabla}_r \times \left[\vec{e}_z \int_{-R/2}^{R/2}dz' \frac{1}{\sqrt{r^2+(z-z')^2}}\right].
\ee
 where $\vec{e}_z$ is the unit vector in the $z$ direction. This expression gives the result
\be\label{bc}
B^C_\theta (r,z) = -\frac{e}{4\pi r}\left(\frac{z-R/2}{\sqrt{r^2+(z-R/2)^2}}-\frac{z+R/2}{\sqrt{r^2+(z+R/2)^2}}\right).
\ee

Substituting for the string term $\bar{E}^{C_{st}}_i= E^C_i +  (\nabla \times B^C)_i$ into the energy functional (\ref{energy}) we get 

\be\label{energy2}
E = \int d^3x \left[\frac{1}{2}\left(-\vec{\nabla}\times\vec{B}+E^C_i +  \vec{\nabla} \times \vec{B}^C\right)^2+\frac{1}{2}(D_i \psi)(D^i \psi)^*+\frac{\lambda}{2}(|\psi|^2-v^2)^2\right].
\ee

Since $E^C_i$ is a perfect gradient, the mixed term in the expansion of the quadratic bracket vanishes so that the field $(\vec{E}^C)^2$ contributes a Coulomb term to the energy:
\be
\int d^3x \frac{1}{2}(\vec{E}^C)^2 = -\frac{e^2}{4\pi R} + \text{(terms independent of $R$)}.
\ee
Neglecting these self-interaction terms we can therefore write the energy as 

 \be\label{energy3}
E = -\frac{e^2}{4\pi R}+\int d^3x \left[\frac{1}{2}\left(-\vec{\nabla}\times\vec{B}+  \vec{\nabla} \times \vec{B}^C\right)^2+\frac{1}{2}(D_i \psi)(D^i \psi)^*+\frac{\lambda}{2}(|\psi|^2-v^2)^2\right].
\ee
with $\vec{B}^C$ given by (\ref{bc}). This energy functional is then minimized with respect to the fields $B_i$ and $\psi$ in order to find the monopole-vortex complex. The nice property of this functional is that it contains, through the presence of the $\vec{B}^C$ term, the effects of the electric monopole sources. \newline

Gauging away the phase of the scalar field 
\be
\psi = S(r,z) e^{i g e \frac{\Omega}{4\pi}}, \quad B_i \rightarrow B_i + \frac{e}{4\pi} \vec{\nabla}\Omega,
\ee
with $S$ real, and picking an ansatz for the gauge field of the form
\be
\vec{B}(\vec{r})=\vec{e}_\theta B(r,z) 
\ee 
we arrive at the energy minimising equations for the field profiles of the form
\be
\partial_z^2 (B-B^C)+\partial_r\left(\frac{1}{r}\partial_r(r(B-B^C))\right)-g^2S^2B =0,
\ee
\be
\partial_z^2S+\frac{1}{r}\partial_r(r \partial_r S)-2\lambda(S^2-v^2)S-g^2B^2S=0.
\ee

The electric field from the string is then given by $\vec{E} =-\partial_z B \;\hat{r} + \frac{1}{r}\partial_r(rB) \;\hat{z}$. Note that the laplacian factor of $B^C$ effectively acts as a delta function, centered on the monopole pair. When $S\rightarrow v$ and gauge symmetry is broken, the field $B$ acquires a mass and decays exponentially.

\section{Non - Abelian Monopole - Vortex complex (NAMVC)}

Now we add orientational moduli to this system in order to promote the system to a non-Abelian monopole-vortex complex. Therefore we consider the action 
\be
S = S_{A}+S_{\chi},
\ee
where $S_A$ is given by equation (\ref{eq1}) and
\be
S_\chi =\int d^4x \left[\frac{1}{2}\partial_\mu\chi^i\partial^\mu \chi_i -\gamma\left((-\mu^2 +|\psi|^2)\chi^2 + \beta \chi^4\right) \right],
\ee
where $\chi ^2 = \chi_i \chi^i$ and $\chi^4 = (\chi_i \chi^i)^2$. In the above $\chi_i$ is a real $SO(3)$ vector, uncharged under the gauge group. 

From here on we switch to the following dimensionless combinations of units
\be
b = \frac{\gamma}{4\lambda} \frac{c-1}{c} , \quad c = v^2/\mu^2, \quad \rho = v r , \quad \tilde{z} = v z,
\ee
we also rescale the fields as
\be
\chi^i = \sqrt{\frac{\mu^2}{2\beta}}\chi.
\ee
\be
S \rightarrow v S \quad B \rightarrow v B.
\ee
We will drop the tildes from our equations, but the reader should be aware that all untilded factors are considered dimensionless in these units.

There is an interesting range of vacua in this model, the vacuum equations $\partial_\psi V = 0$ and $\partial_\chi V=0$ lead to several branches of solutions, which are
\bea\label{vacua}
S^2= 0, \quad \chi^2 = 1,\\
S^2 = 1, \quad \chi^2 =0, \\
S^2 =\frac{b-4(c-1)\beta\lambda}{bc-4(c-1)\beta\lambda},\quad \chi^2 = \frac{4(c-1)^2\beta\lambda}{bc-4(c-1)\beta\lambda},
\eea
with the second and the third branch coalescing at the special point $c =1$.  These vacua represent different physics. The first vacuum breaks the global symmetry  carried by $\chi$ but not the gauge symmetry, the second vacuum does the opposite and the third breaks both. For non-Abelian vortex solutions we will be interested in the second vacuum, however we will also investigate the first. In the above dimensionless units the vacuum stability requirement for vacuum II is
\be
\beta > \frac{b}{4\lambda c(c-1)}.
\ee

It turns out that for a stable vacuum the third of these solutions always leads to a complex $\chi$ field, hence we ignore it from now on.

\subsection{Uniform NAMVC - Vacuum II}

Consider first an ansatz of the form

\be
\chi^i = \sqrt{\frac{\mu^2}{2\beta}}\chi(\rho,z )(0,0,1),
\ee
\be
S= S(\rho, z), \quad B= B(\rho,z).
\ee

This ansatz is particularly simple as it forces the $\chi$ field to point in a single direction of isospace, leaving a $U(1)$ subgroup of the original $SO(3)$ group invariant. With this ansatz the equations of motion of the coupled system reduce to (we use the shifted field  $\tilde{B} = B-B^c$ and drop all tildes on $z$)
\be
\partial_{z}^2\tilde{B}+\partial_\rho\left(\frac{1}{\rho}\partial_\rho(\rho\tilde{B})\right)-g^2S^2(\tilde{B}+B^C) =0,
\ee
\be
\partial_{z}^2S+\frac{1}{\rho}\partial_\rho(\rho \partial_\rho S)-2\lambda(S^2-1)S-g^2(\tilde{B}+B^C)^2S-\frac{b}{2\beta(c-1)}\chi^2 S=0,
\ee
\be
\partial_{z}^2\chi+\frac{1}{\rho}\partial_\rho(\rho \partial_\rho \chi)-\frac{b}{c-1} (c S^2+\chi^2-1)\chi=0.
\ee

We intend to solve these equations with the following boundary conditions
\be
S(0, m \leftrightarrow m) = 0, \quad S(0 , \pm \infty) = 1, 
\ee
\be
\partial_r\chi(0, m \leftrightarrow m) = 0, \quad \chi(0 , \pm \infty) = 0, 
\ee
\be
\tilde{B}(0, m \leftrightarrow m) = 0, \quad \tilde{B}(0, \pm \infty) = 0, 
\ee
\be
S(\infty, z) = 1, \quad \chi(\infty, z) = 0, \quad \tilde{B}(\infty ,  z) = 0.
\ee

\begin{figure}[ptb]
\begin{subfigure}{.5\textwidth}
\centering
\includegraphics[width=0.9\linewidth]{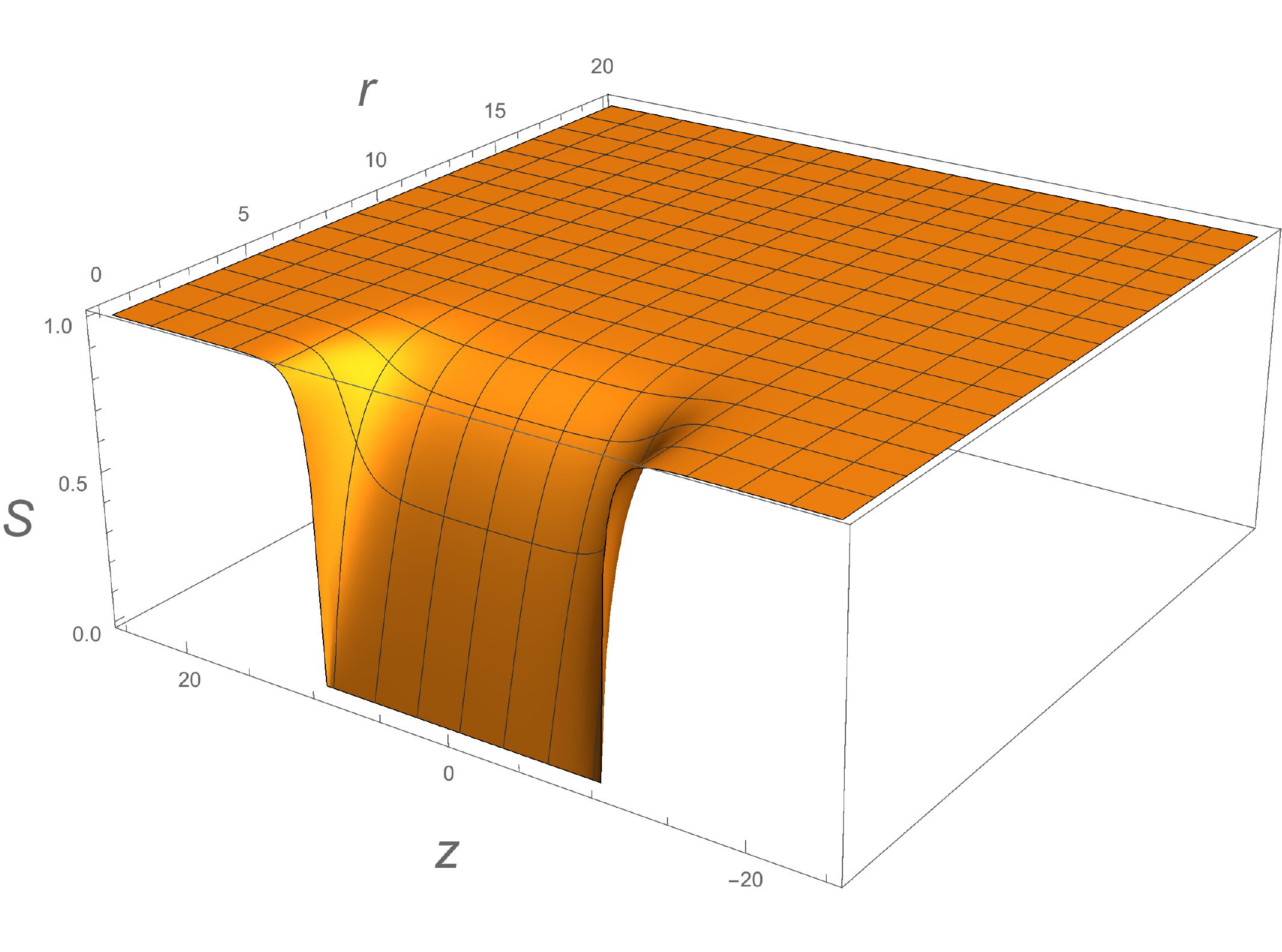}
\caption{$S$}
\end{subfigure}
\begin{subfigure}{.5\textwidth}
\centering
\includegraphics[width=0.9\linewidth]{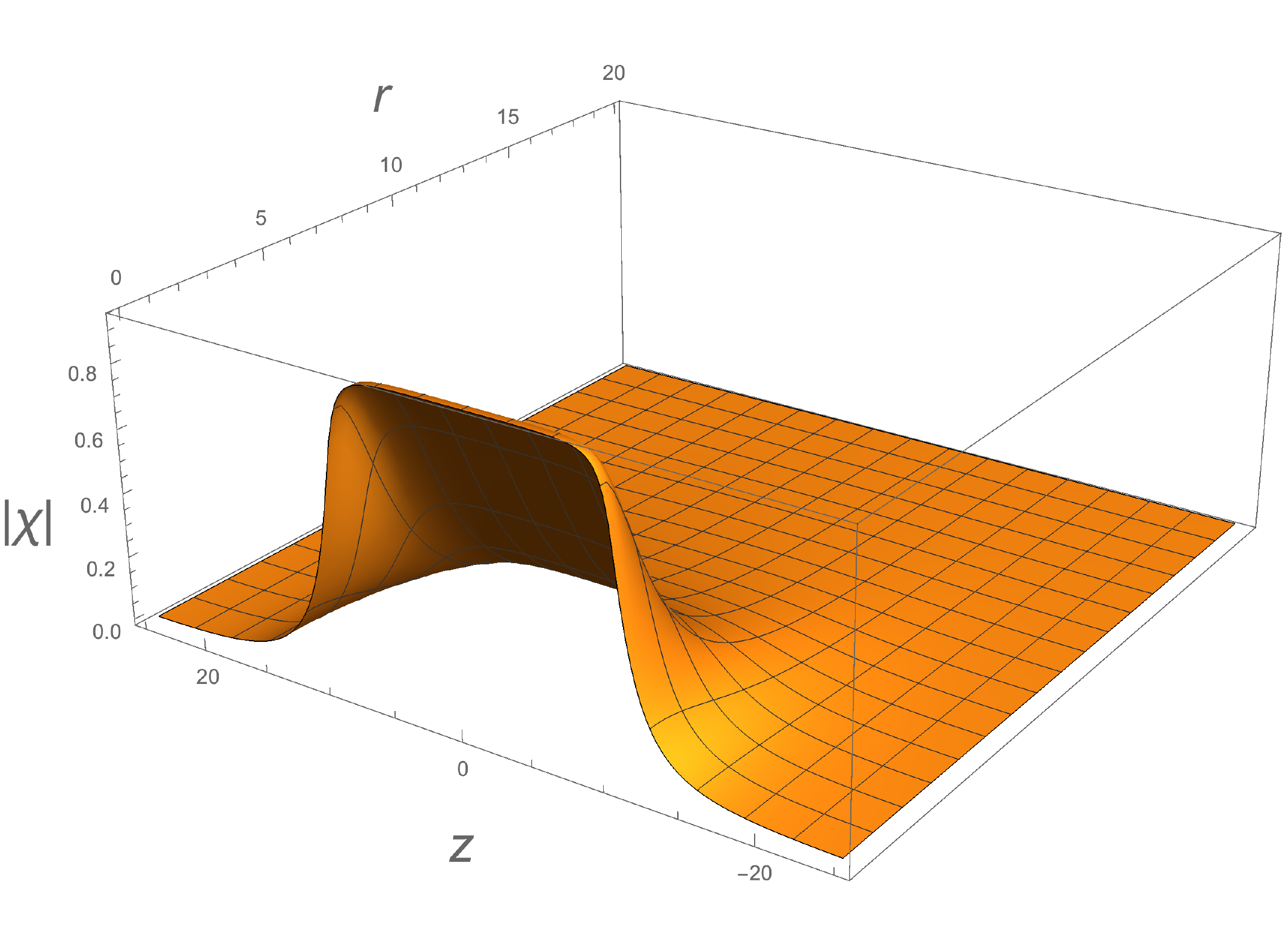}
\caption{$\chi$}
\end{subfigure}
\begin{subfigure}{0.5\textwidth}
\centering
\includegraphics[width=0.8\linewidth]{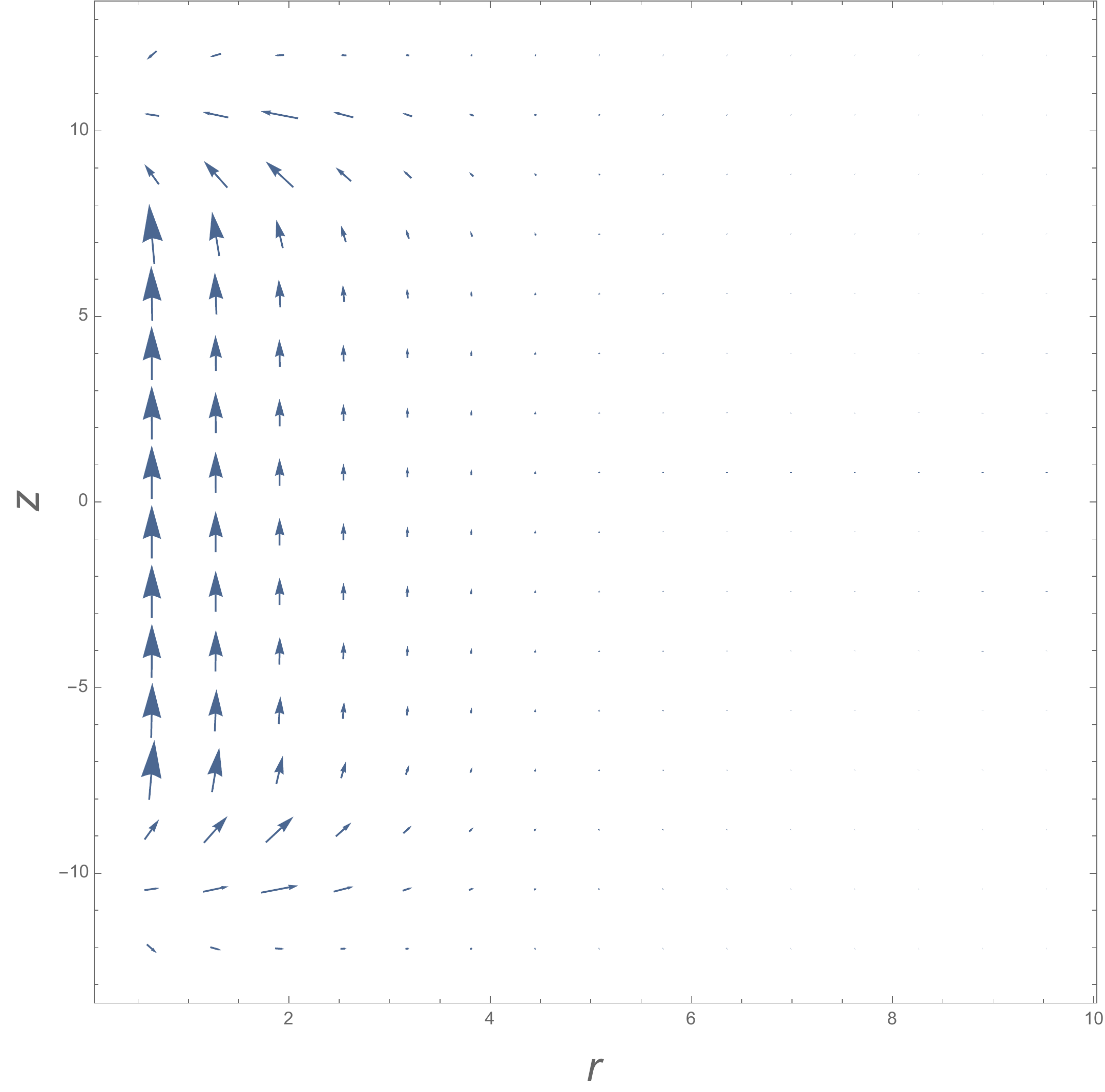}
\caption{$E$}
\end{subfigure}
\begin{subfigure}{0.5\textwidth}
\centering
\includegraphics[width=0.6\linewidth]{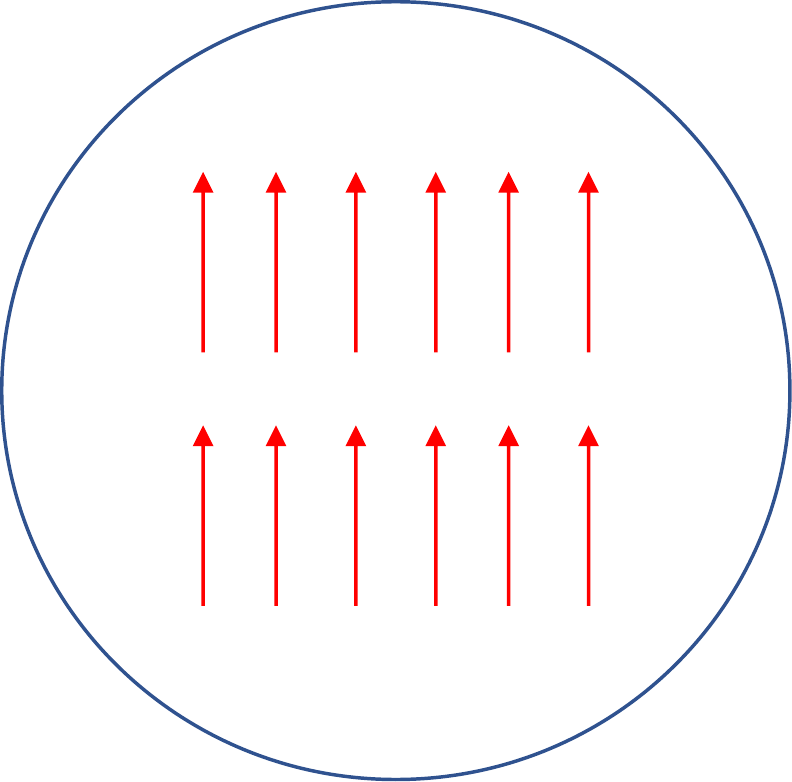}
\caption{}
\end{subfigure}
\caption{Field profiles for the non-Abelian Monopole Vortex complex at $g = 0.8$, $e=2\pi /g$,  $\lambda = 0.42$, $c = 1.01$, $b=0.05$, $\beta = 5.45$, $R = 20$. Plot (c) shows the electric field distribution derived from $B$. Figure d) shows a diagram of the isospace orientation of the $\chi$ field in a cross section of the flux tube.}%
\label{fig1}%
\end{figure}

\begin{figure}[ptb]
\centering
\includegraphics[width=0.5\linewidth]{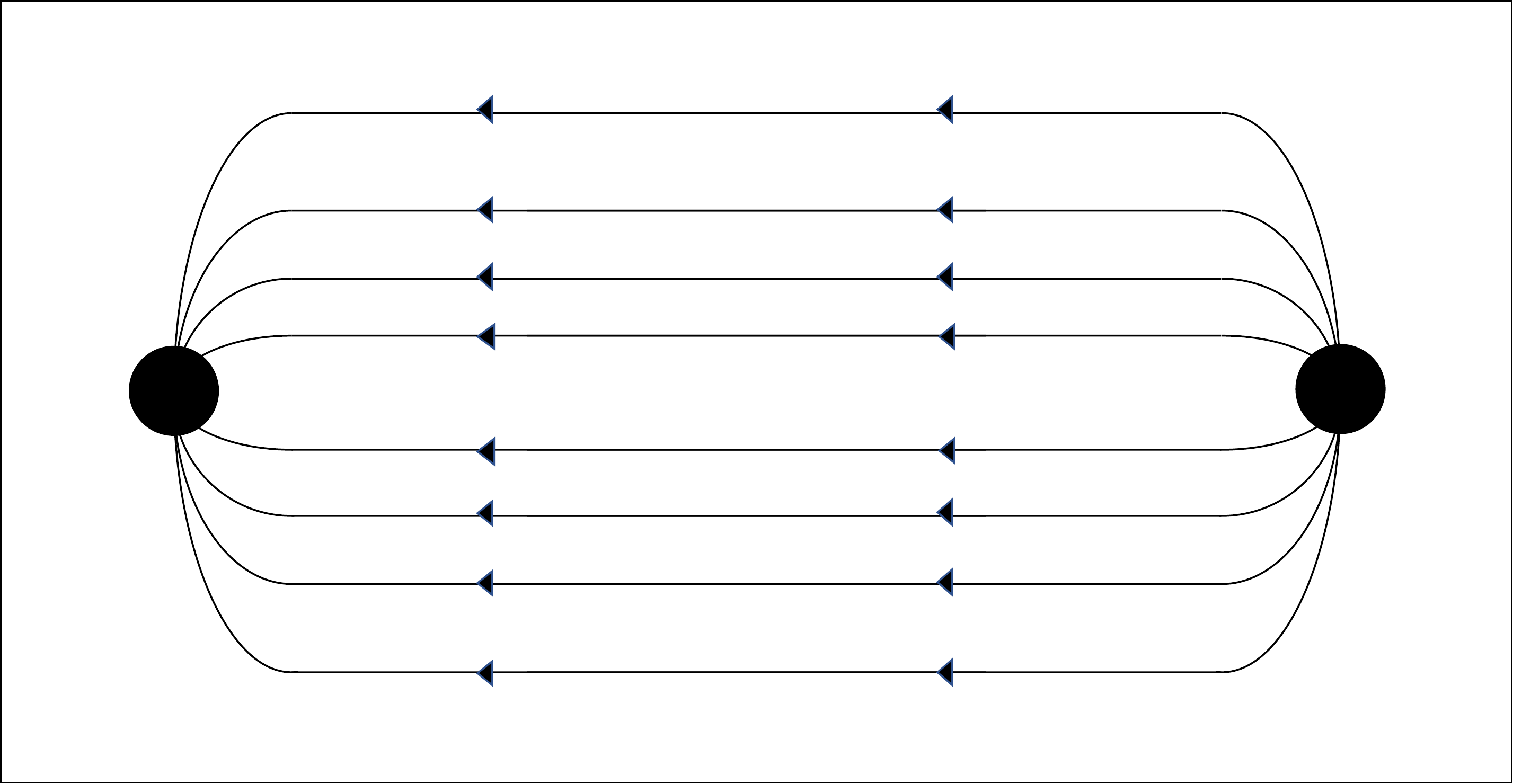}
\caption{Schematic diagram of flux confinement for solution presented in figure \ref{fig1}. The rectangle denotes the superconducting region.}%
\label{figconf2}%
\end{figure}

In the above $m \leftrightarrow m$ denotes the region of $z$ between the monopole sources. We solve the equations numerically using a relaxation procedure, the order of accuracy of our solutions is $\mathcal{O}(10^{-6})$. The solutions are shown in figure \ref{fig1}. A schematic representation of the flux is shown in figure \ref{figconf2}.

These solutions represent a confining non-Abelian electric flux tube between the monopole charges. We will analyze its properties further in later sections.

\subsection{Hedgehog MVC and superconducting droplets - Vacuum I}

The presence of the $\chi$ field in this system generates a different kind of interesting solution. There is another natural ansatz for the $\chi^i$ vector which is
\be
\chi^i = \chi_\rho(\rho, z)(\cos\theta,\sin\theta,0)+\chi_z(\rho,z)(0,0,1).
\ee

The reader will notice this as a hedgehog like structure, with the internal space vector pointing in a radial direction at far distances (by radial here we mean the radial direction of a large sphere).  For this kind of configuration to be able to form a stable solution we need the $\chi$ field to condense in the vacuum and hence protect the hedgehog topologically. We must therefore consider the first of the vacua shown in equation (\ref{vacua}). This vacuum is the true vacuum when
\be
\beta < \frac{b}{4\lambda c(c-1)}.
\ee
In this vacuum, the gauge symmetry is not broken but the internal global symmetry carried by $\chi$ is. These solutions will represent finite regions of superconductivity (where $S$ is non-zero), commonly known as superconducting droplets. The purpose of this section is therefore to investigate what happens when these droplets are subjected to the flux of a monopole anti-monopole pair. Since the $\chi$ field is ungauged and condenses in the vacuum, the energy is infrared divergent and these solutions must be considered in a finite box.  \\

With the above ansatz for the $\chi$ field (all other fields unchanged) the equations of motion reduce to

\be
\partial_{z}^2\tilde{B}+\partial_\rho\left(\frac{1}{\rho}\partial_\rho(\rho\tilde{B})\right)-g^2S^2(\tilde{B}+B^C) =0,
\ee
\be
\partial_{z}^2S+\frac{1}{\rho}\partial_\rho(\rho \partial_\rho S)-2\lambda(S^2-1)S-g^2(\tilde{B}+B^C)^2S-\frac{b}{2\beta(c-1)}(\chi_\rho^2+\chi_z^2) S=0,
\ee
\be
\partial_{z}^2\chi_z+\frac{1}{\rho}\partial_\rho(\rho \partial_\rho \chi_z)-\frac{b}{c-1} (c S^2+\chi_\rho^2+\chi_z^2-1)\chi_z=0.
\ee
\be
\partial_{z}^2\chi_\rho+\frac{1}{\rho}\partial_\rho(\rho \partial_\rho \chi_\rho)-\frac{\chi_\rho^2}{\rho^2}-\frac{b}{c-1} (c S^2+\chi_\rho^2+\chi_z^2-1)\chi_\rho=0.
\ee

We solve these with the boundary conditions
\be
S'(0, m \leftrightarrow m) = 0, \quad S(0 , \pm \infty) = 0, 
\ee
\be
\partial_\rho\chi_\rho(0, z) = 0,  \quad \chi_\rho(\infty, z) \rightarrow \rho/\sqrt{\rho^2+z^2},
\ee
\be
\partial_\rho\chi_z(0, z) = 0,  \quad \chi_z(\infty, z) \rightarrow z/\sqrt{\rho^2+z^2},
\ee
\be
\tilde{B}(0, m \leftrightarrow m) = 0, \quad \tilde{B}(0, \pm \infty) = 0, 
\ee
\be
S(\infty, z) = 0, \quad \chi(\infty, z) = 0, \quad \tilde{B}(\infty ,  z) = 0.
\ee

\begin{figure}[ptb]
\begin{subfigure}{.5\textwidth}
\centering
\includegraphics[width=0.9\linewidth]{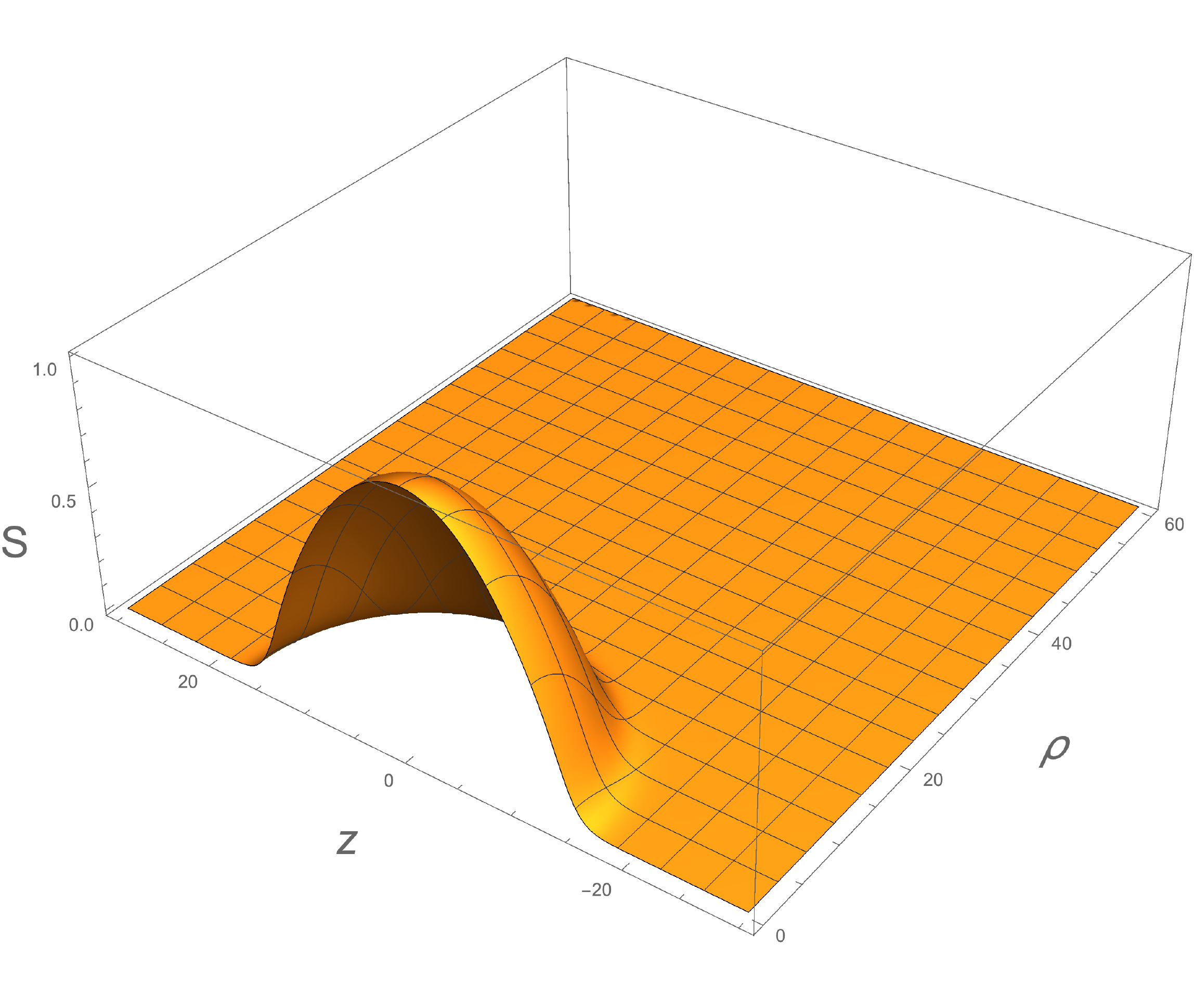}
\caption{$S$}
\end{subfigure}
\begin{subfigure}{.5\textwidth}
\centering
\includegraphics[width=0.9\linewidth]{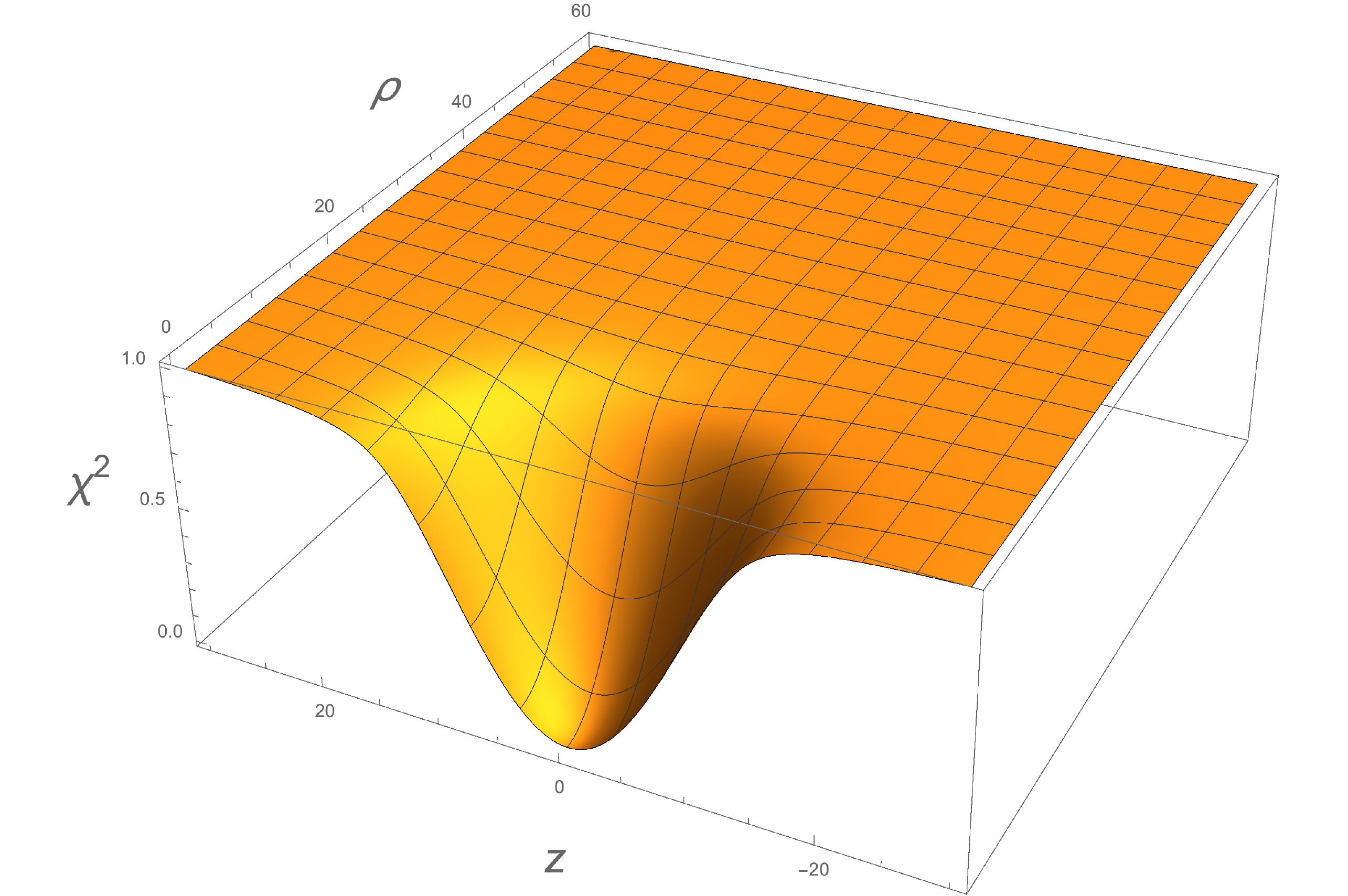}
\caption{$\chi_\rho^2+\chi_z^2$}
\end{subfigure}
\begin{subfigure}{0.5\textwidth}
\centering
\includegraphics[width=0.8\linewidth]{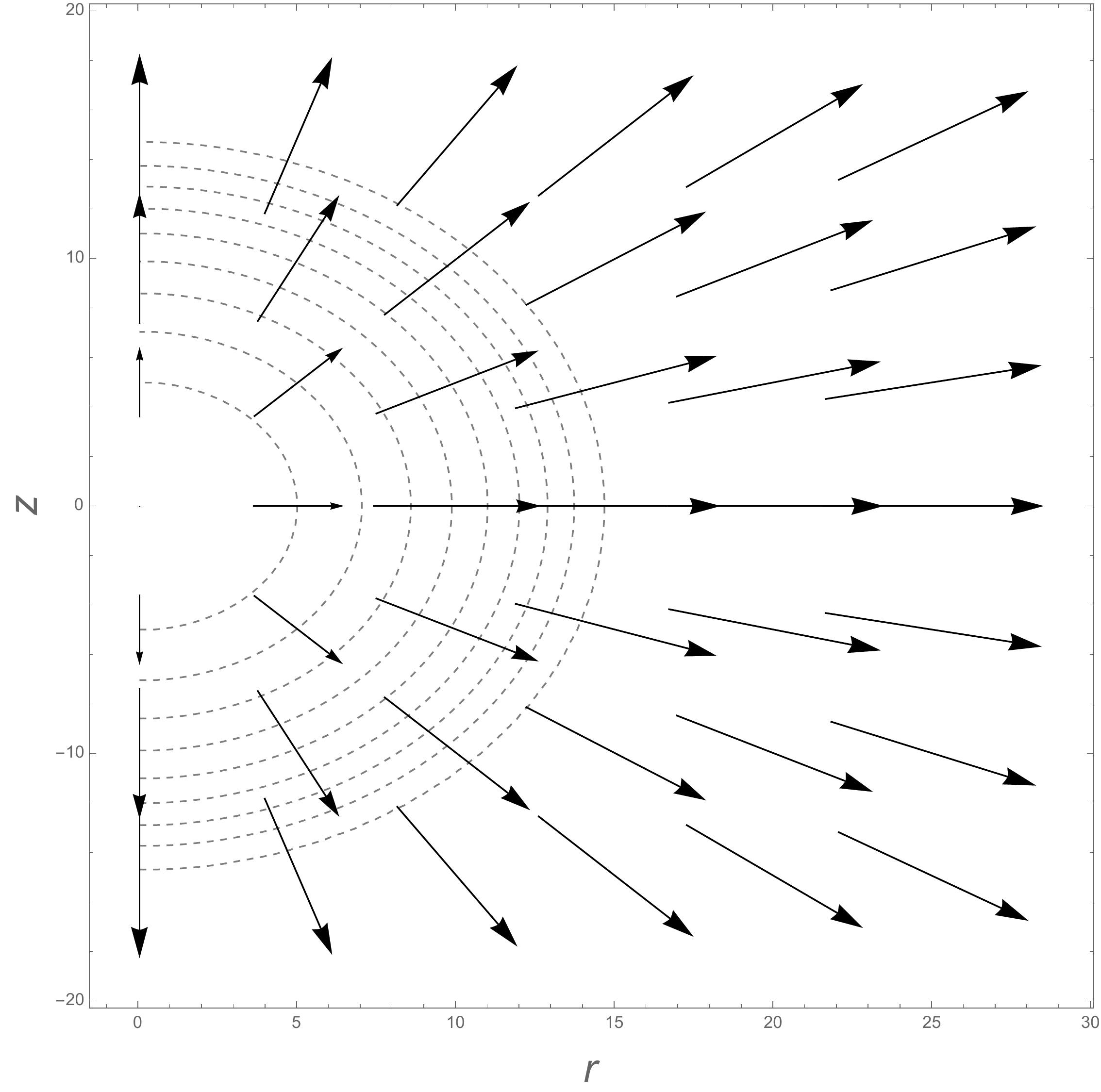}
\caption{$\vec\chi$}
\end{subfigure}
\begin{subfigure}{0.5\textwidth}
\centering
\includegraphics[width=0.6\linewidth]{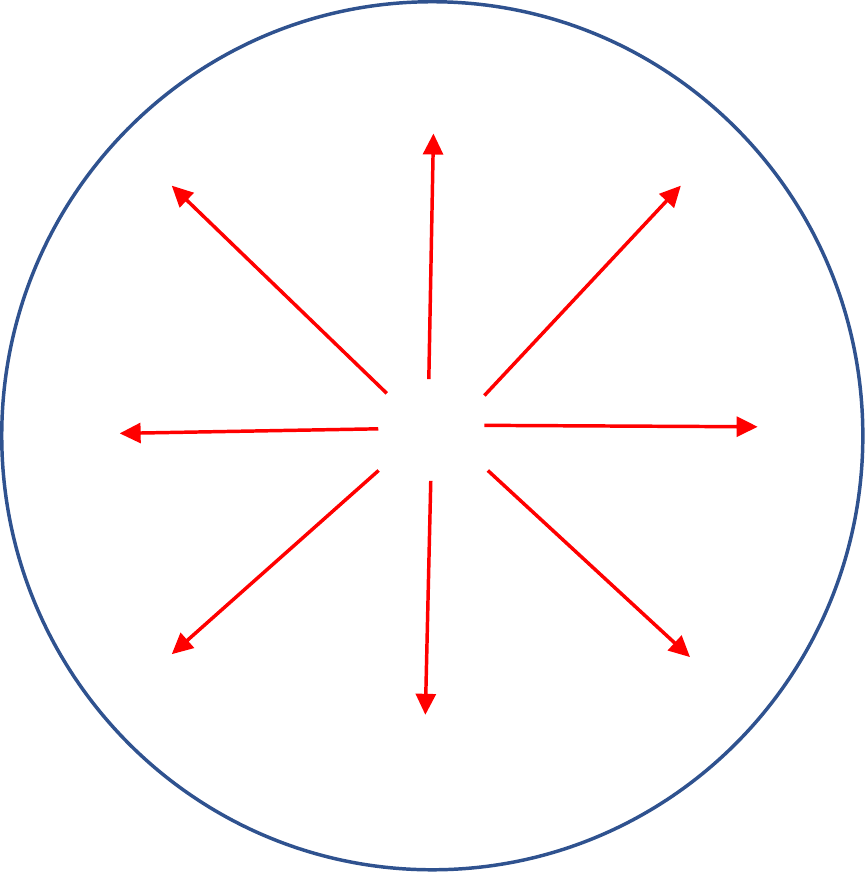}
\caption{}
\end{subfigure}
\caption{Field profiles for the non-Abelian Monopole Vortex complex at $g = 0.8$, $e=0$,  $\lambda = 10$, $c = 1.05$, $b=0.005$, $\beta = 0.9 b/(4\lambda c(c-1))$. Plot (c) shows the chi vector plot on top of the droplet. Figure d) shows a schematic diagram of the isospace orientation of the $\chi$ field in a cross section of the droplet at $z=0$.}%
\label{fig13}%
\end{figure}

\begin{figure}[ptb]
\begin{subfigure}{.5\textwidth}
\centering
\includegraphics[width=0.9\linewidth]{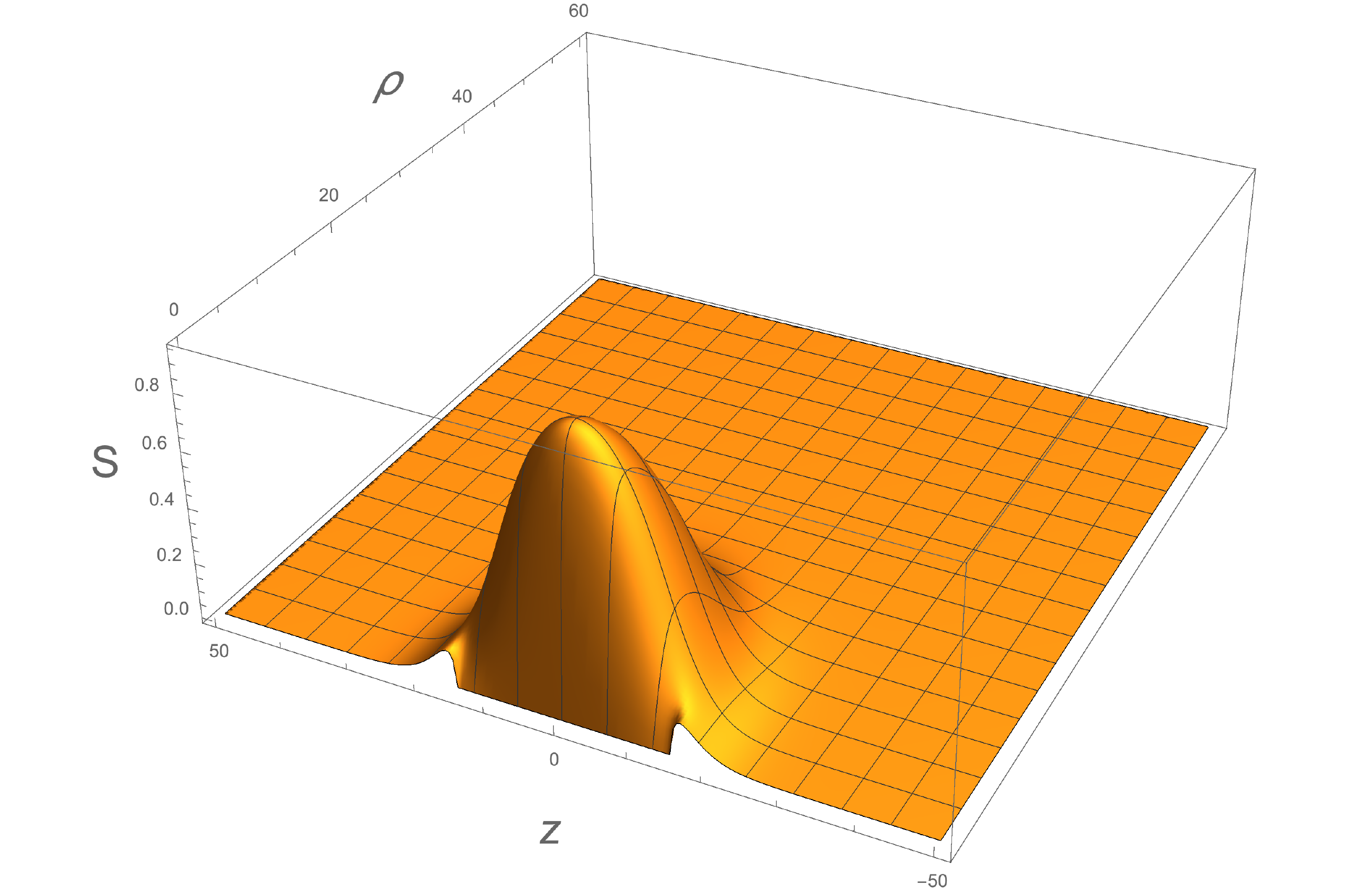}
\caption{$S$}
\end{subfigure}
\begin{subfigure}{.5\textwidth}
\centering
\includegraphics[width=0.9\linewidth]{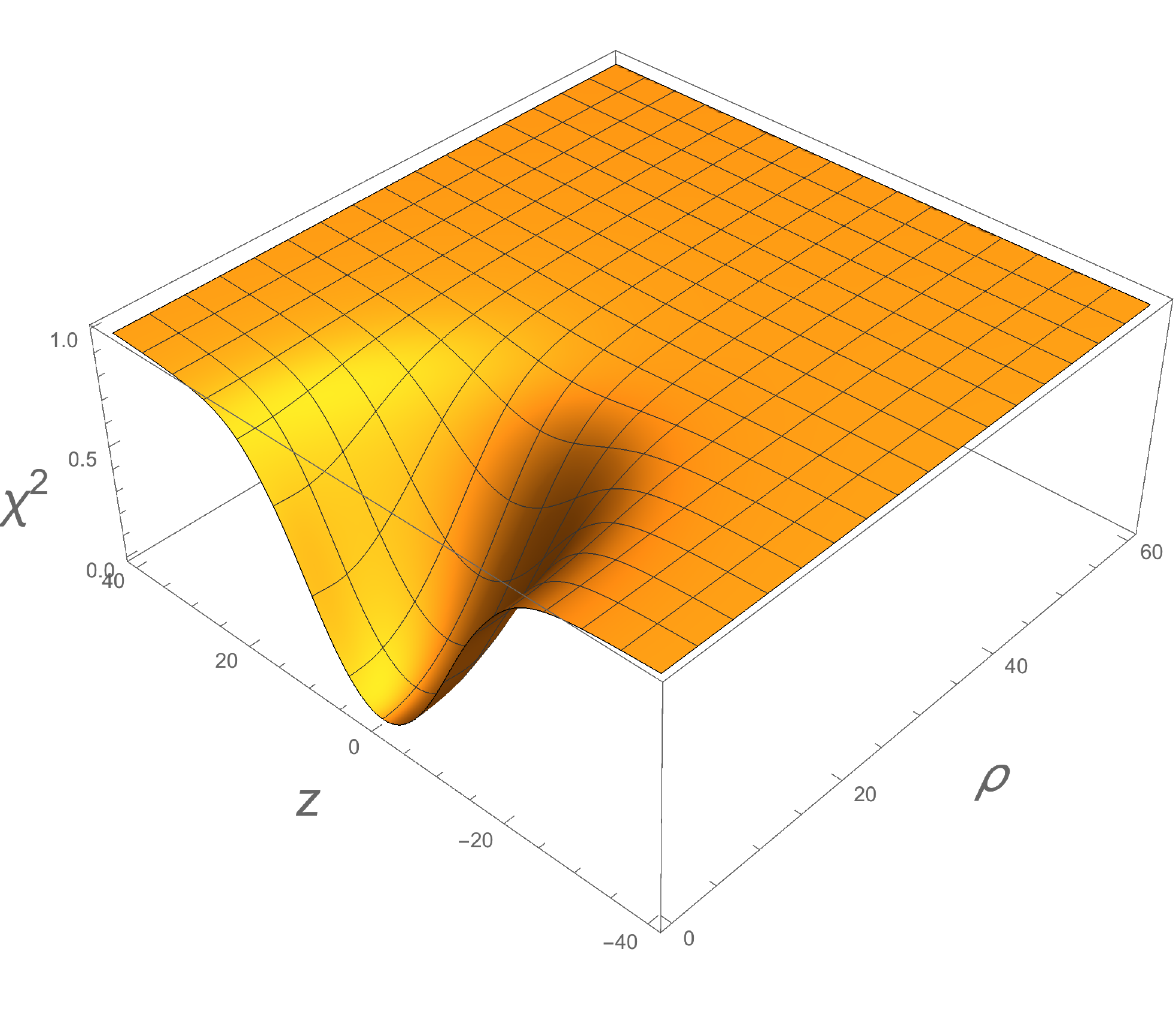}
\caption{$\chi_\rho^2+\chi_z^2$}
\end{subfigure}
\begin{subfigure}{0.5\textwidth}
\centering
\includegraphics[width=0.8\linewidth]{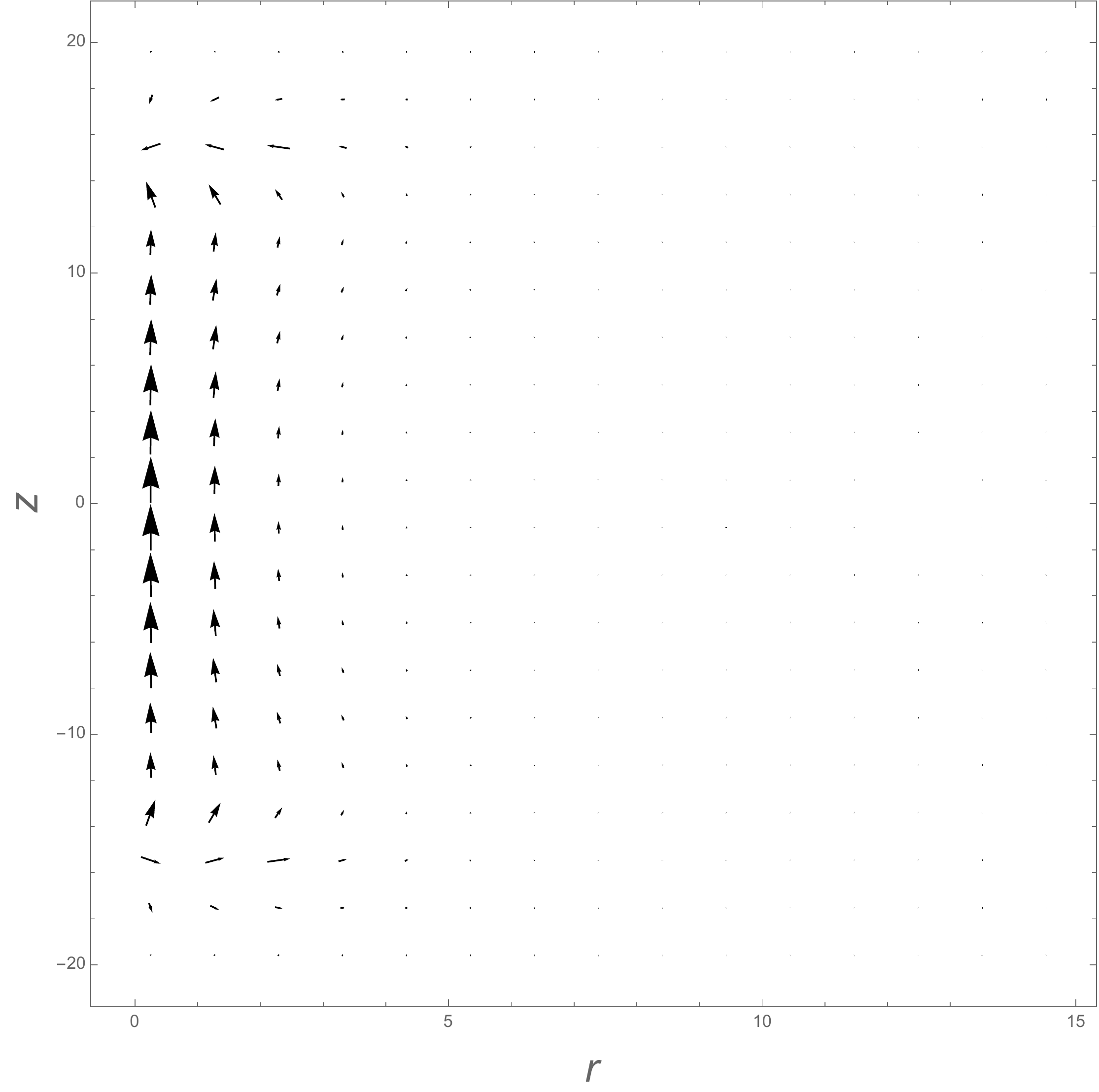}
\caption{$\vec E$}
\end{subfigure}
\begin{subfigure}{0.5\textwidth}
\centering
\includegraphics[width=0.8\linewidth]{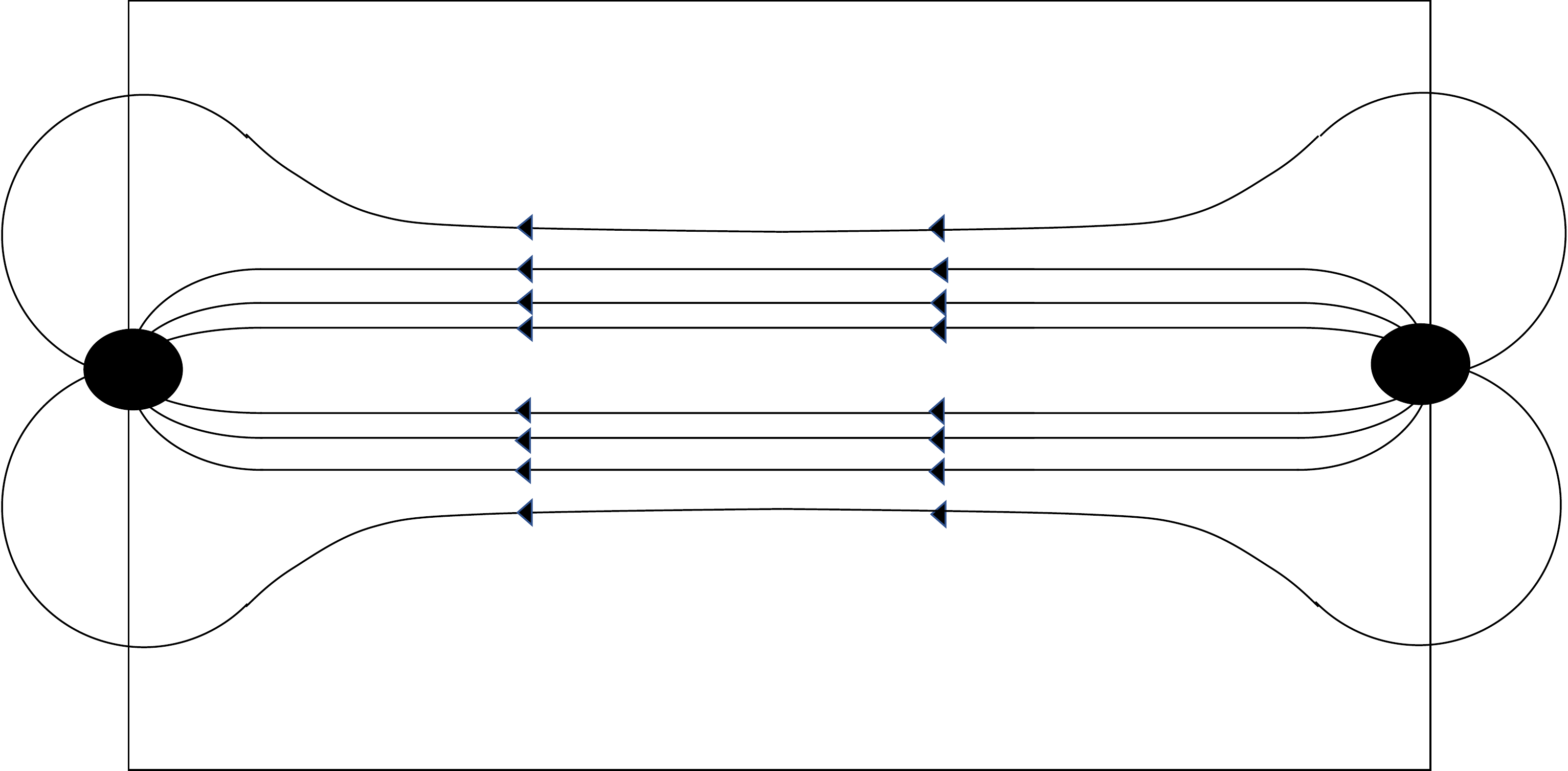}
\caption{}
\end{subfigure}
\caption{Field profiles for the non-Abelian Monopole Vortex complex contained inside the droplet at $g = 0.8$, $e=2\pi/g$,  $\lambda = 0.5$, $c = 1.05$, $b=0.003$, $\beta = 0.9 b/(4\lambda c(c-1))$, $R=30$. Plot (c) shows the electric field vector plot inside the droplet. This case is similar to the previous NAMVC case. Plot d) shows a schematic of the flux confinement for this solution.}%
\label{fig11}%
\end{figure}

In figure \ref{fig13} we present one such solution in which the external monopoles are switched off, i.e. $e=0$. This is representative of the of a superconducting droplet. If we turn the field on, then we must solve with the condition that
\be
S(0, m \leftrightarrow m) = 0,
\ee
all other conditions unchanged, since the divergence of $B^C$ at $\rho=0$ forces $S$ to be zero between the monopoles. There are two interesting cases when the flux is turned on: when the monopole pair is fully contained within the superconducting droplet, or when it is outside. Figure \ref{fig11} shows the solutions in which the monopole pair is placed within the superconducting droplet, as expected the electric flux penetrates the superconducting region and confines the monopole anti-monopole pair. This case is similar to the previous case shown in figure \ref{fig1}. There is a subtle difference in the vector plot close to monopole sources. The curvature of the electric field is slightly outgoing, this is a boundary effect. Since the monopole pair is close to the end points of the superconducting droplet on the $z$-axis the flux is free to escape when this ends, which is why it wants to curve outwards. A schematic digram of this is shown in figure \ref{fig11} (d). \\

\begin{figure}[ptb]
\begin{subfigure}{.5\textwidth}
\centering
\includegraphics[width=0.9\linewidth]{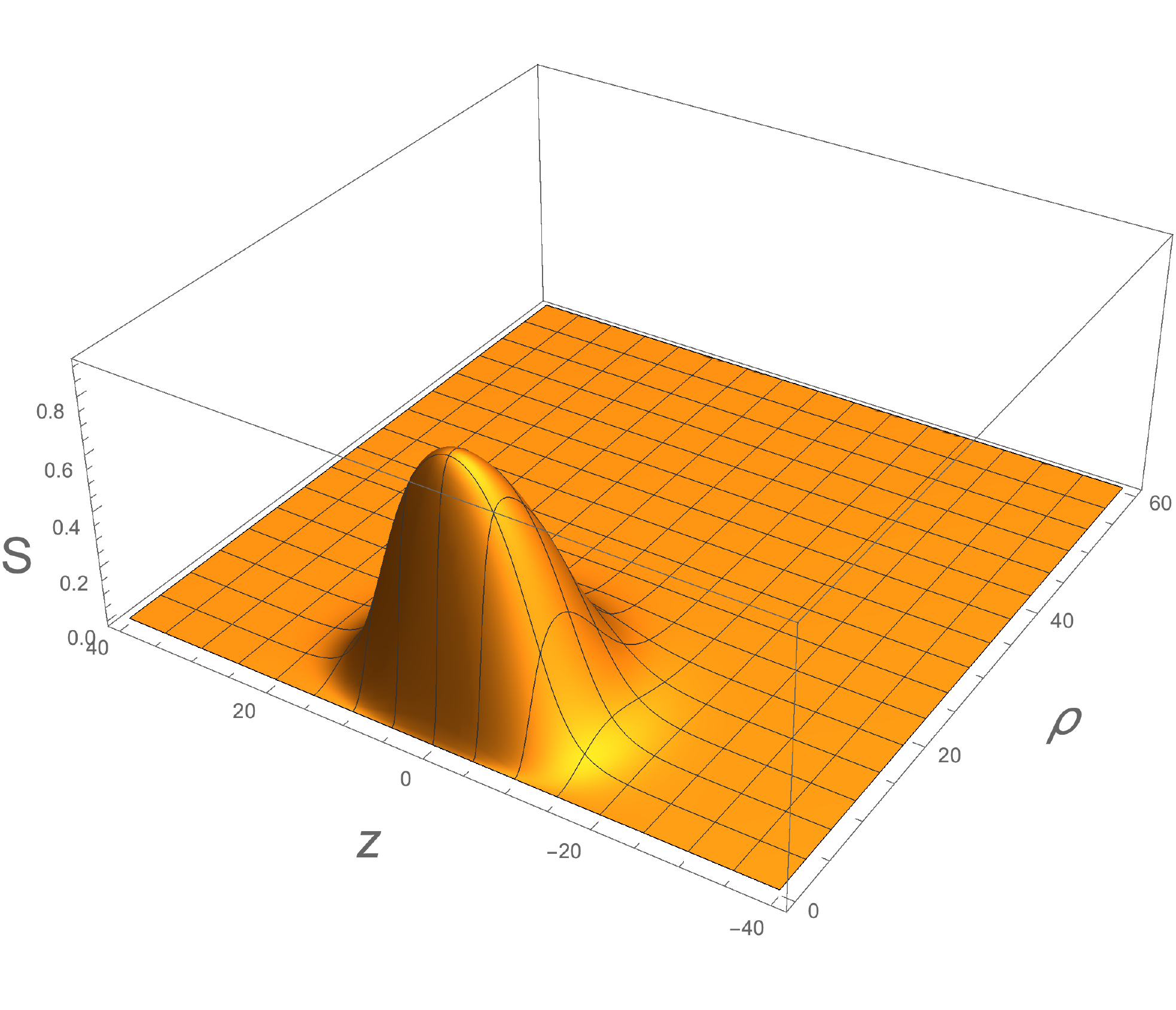}
\caption{$S$}
\end{subfigure}
\begin{subfigure}{.5\textwidth}
\centering
\includegraphics[width=0.9\linewidth]{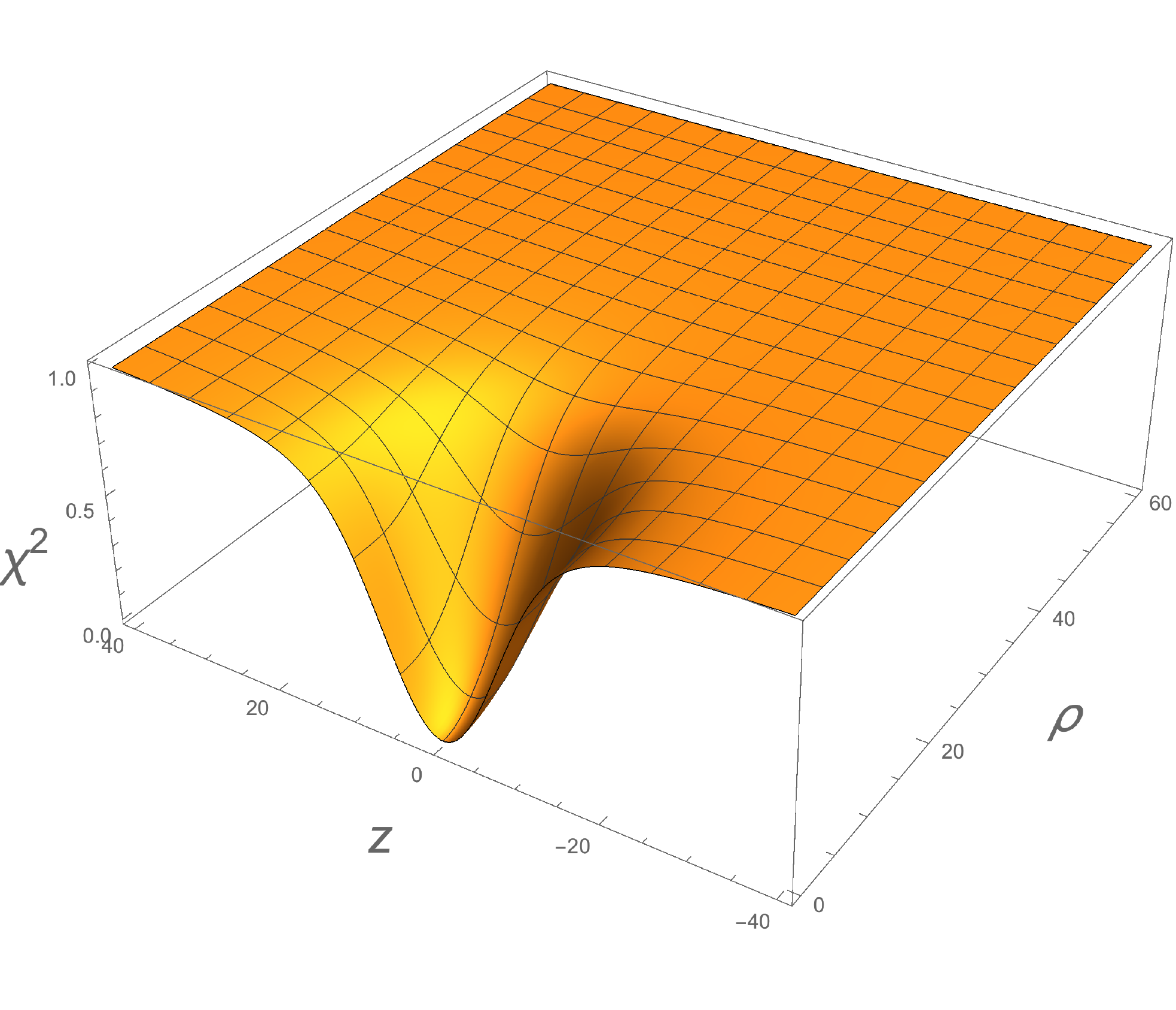}
\caption{$\chi_\rho^2+\chi_z^2$}
\end{subfigure}
\begin{subfigure}{0.5\textwidth}
\centering
\includegraphics[width=0.8\linewidth]{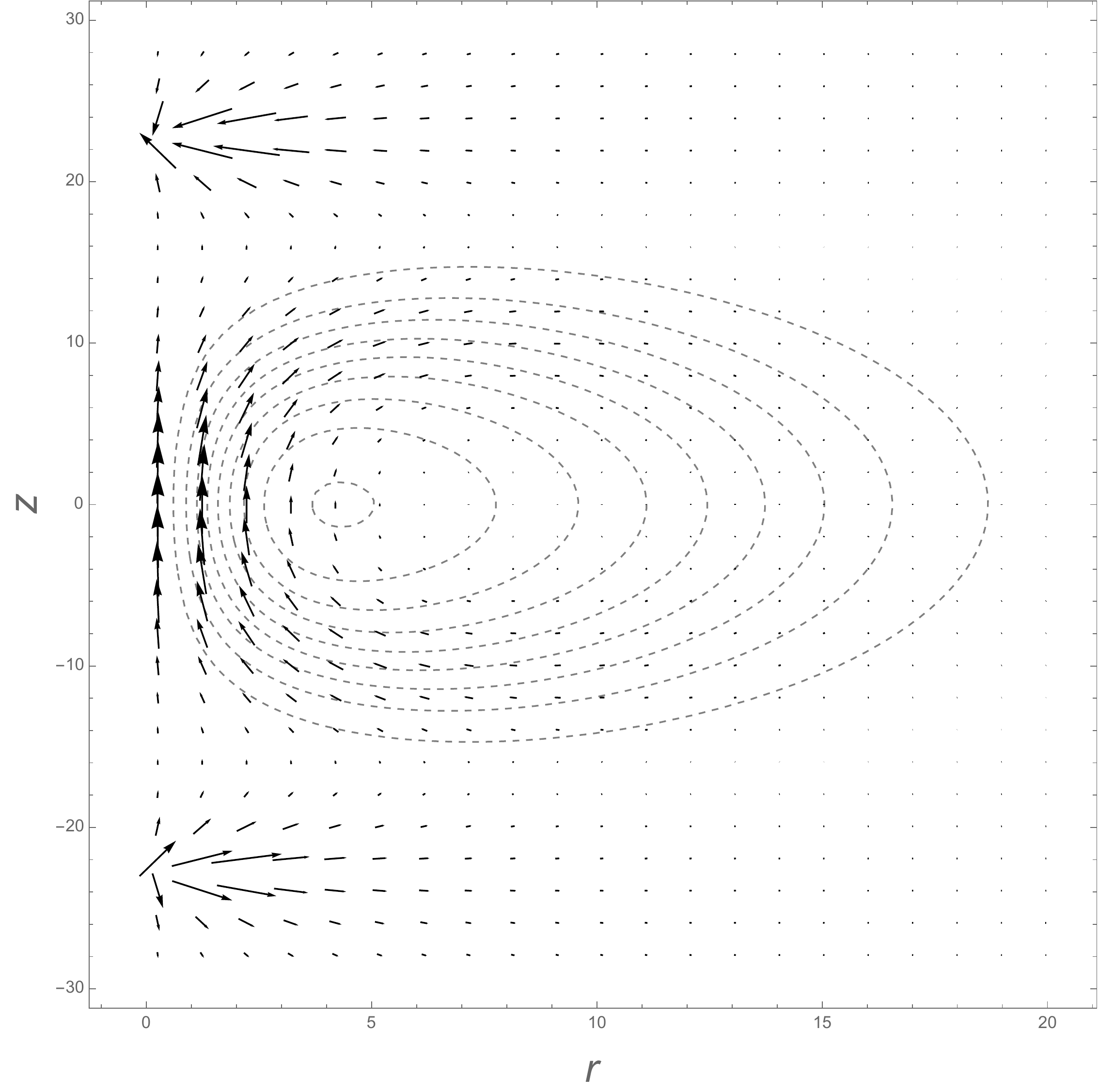}
\caption{$\vec E$}
\end{subfigure}
\begin{subfigure}{0.5\textwidth}
\centering
\includegraphics[width=0.8\linewidth]{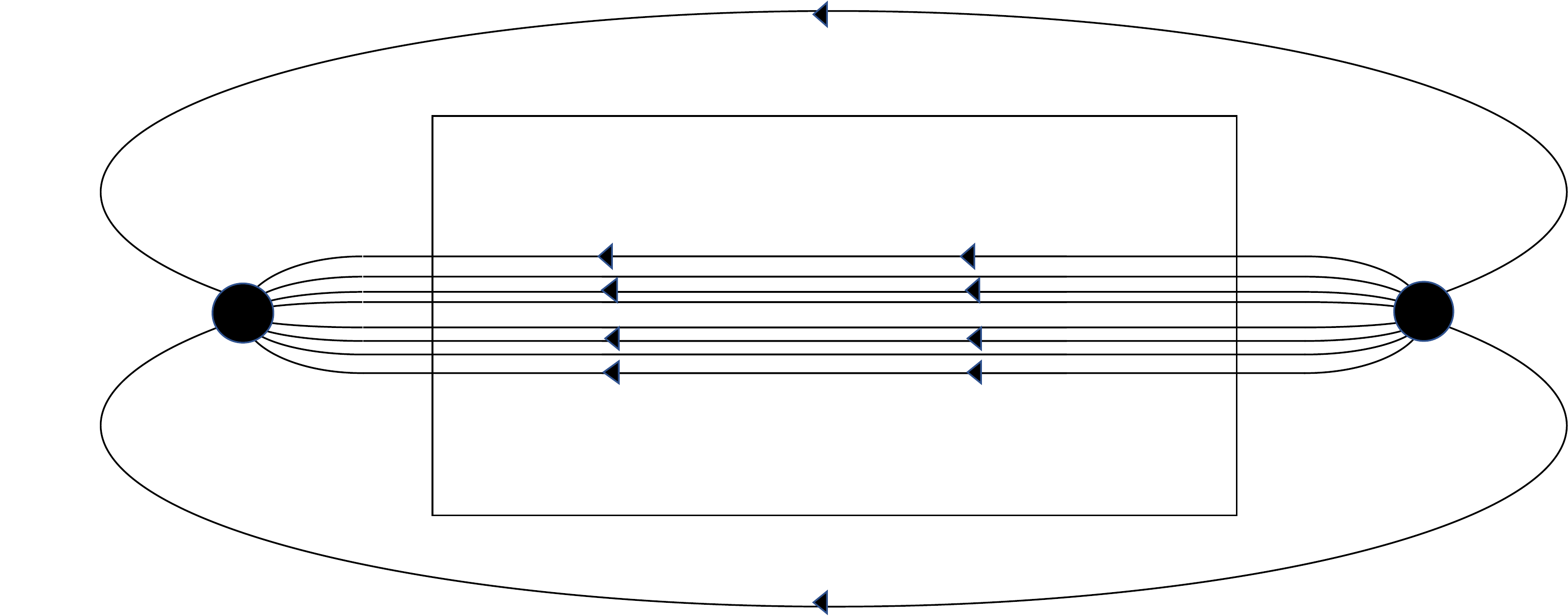}
\caption{}
\end{subfigure}
\caption{Field profiles for the non-Abelian Monopole Vortex complex contained inside the droplet at $g = 0.8$, $e=4\pi/g$,  $\lambda = 0.5$, $c = 1.05$, $b=0.003$, $\beta = 0.9 b/(4\lambda c(c-1))$, $R=45.5$. Plot (c) shows the electric field vector plot over the whole region (we include here the contribution from the monopoles to mark their position) superposed on a contour plot of the $S$ field. Plot d) shows a schematic of the electric flux, some flux for this solution is free to go round the superconducting dipole.}%
\label{fig15}%
\end{figure}

The case in which the monopole pair is placed far from the droplet is shown in figure \ref{fig15}.  Here we see that most of the flux penetrates through the droplet, close to $r=0$, but some is allowed to go around it, unhindered by the effective Meissner effect. This corresponds schematically to the case shown in figure \ref{fig15} (d).

\section{Energetics}

This section is devoted to the analysis of the energy of the NAMVC solution. In particular we must check that the NAMVC is energetically preferred over the MVC with no orientational degrees of freedom. The energy functional of the system is (in a-dimensional units)
\be
E = E_S + E_\chi-\frac{e^2}{4\pi R},
\ee
where
\be\nn
E_S = \int \rho d\rho\; dz  \left[\frac{1}{2}\left(-\vec{\nabla}\times\vec{B}+  \vec{\nabla} \times \vec{B}^C\right)^2+\frac{1}{2}\partial_i S\partial^i S+ \frac{e^2}{2}S^2(B+B_c)^2+\frac{\lambda}{2}(S^2-1)^2\right],
\ee
\be
E_\chi = \int \rho d\rho\; dz \frac{1}{2}\; \left[\frac{1}{2c\beta}(\nabla \chi)^2+\frac{1}{4\beta}\frac{b}{c(c-1)}\left(\frac{1}{2}\chi^4-\chi^2(1-c S^2)\right)\right].
\ee
For the solution shown in figure \ref{fig1} we find that the energy of the MVC without the $\chi$ field is
\be
E_S= 73.5,
\ee
whilst that of the NAMVC is
\be
E_\chi = 65.4.
\ee
This shows that, at least for this set of parameters, the NAMVC is energetically preferred over its abelian counterpart. We have not found a range of parameters where the NAMVC is not the energetically lower solution.  This is enough to say that the NAMVC in this model is metastable at least within the probe quark limit.  \newline

\begin{figure}[ptb]
\centering
\includegraphics[width=0.6\linewidth]{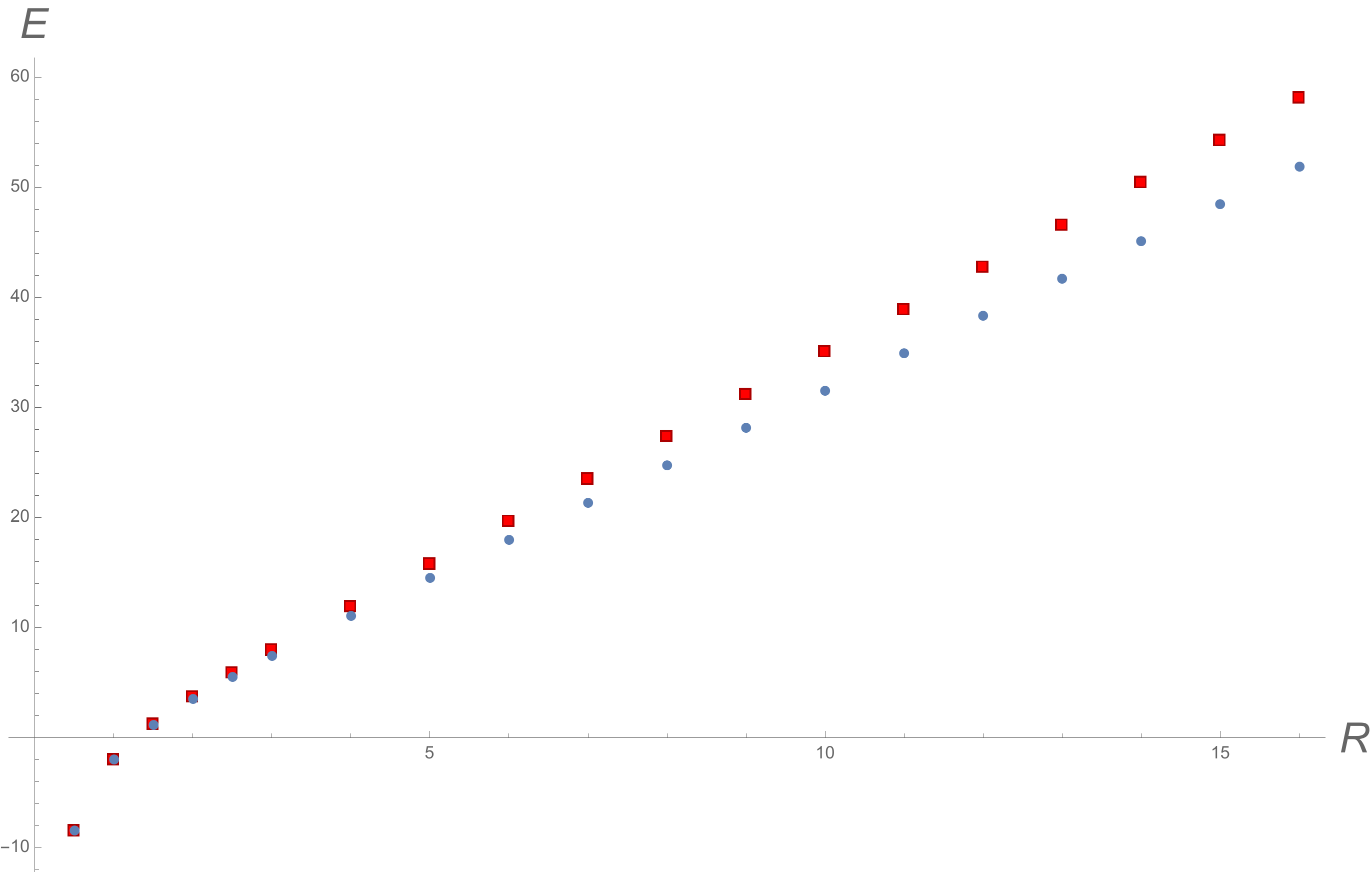}
\caption{Energy dependence on monopole anti-monopole separation $R$. The blue circles represent the NAMVC solution, whilst the red squares the standard MVC.}%
\label{figconf}%
\end{figure}

We are also interested in whether the non-Abelian field influences the confinement properties of the flux tube. For this purpose, we show in figure \ref{figconf} a plot of the energy of our system against the monopole anti-monopole separation $R$. Both the NAMVC and the standard MVC are plotted. The results indicate that the confining property of the soliton, namely linear dependence on $R$, isn't affected. The energy of the $\chi$ configuration being lower means a smaller gradient of the linear sector of this plot, which in turns means that since this field lowers the tension of the string, confinement is weaker. \\

The hedgehog solution, in which the ungauged $\chi$ field condenses in the vacuum is energetically divergent in the infrared. This solution must therefore be regularized by considering it in a finite box. We will not consider the energetics of this solution further.

\section{Low Energy Theory}

This section is devoted to the analysis of the low energy theory living on the monopole - vortex world sheets. That is we mean to uncover and describe the theory of gapless excitations of the soliton solutions found above. Similar but more involved considerations appear in \cite{Cipriani:2012pa} relating to the aforementioned models.

\subsection{The uniform NAMVC}

\begin{figure}[ptb]
\centering
\includegraphics[width=0.6\linewidth]{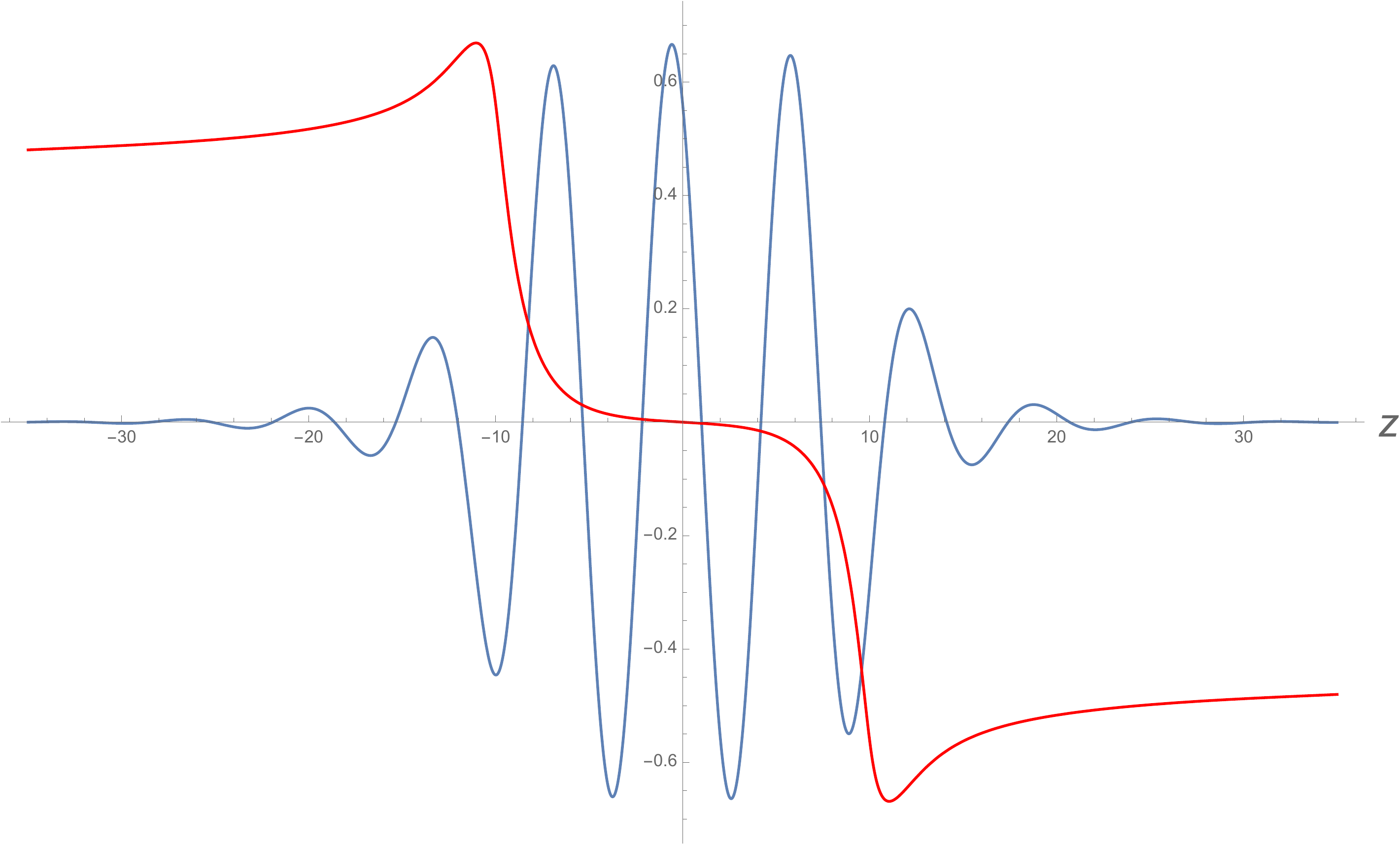}
\caption{Orientational mode perturbation along the $z$ axis. The oscillation, shown in blue, can only occur where $\chi$ is non-vanishing, in between the monopole sources, and is severely damped outside. The potential term is given by the function $g(z)$, shown in red.  $\omega^2 = 0.98$}%
\label{fig3}%
\end{figure}

Firstly, since we consider the monopoles as infinitely massive charge sources there are no translational gapless excitations. The more interesting case are the orientational moduli. The pattern of symmetry breaking in the core $SO(3) \rightarrow U(1)$ means the orientational moduli form a $CP(1)$ non-linear sigma model on the soliton world-sheet. There should therefore be two orientational degrees of freedom. In the case of an infinite string this is easy to recover, one simply replaces 
\be
\chi^i \rightarrow \chi(\rho) S^{i}(z,t),
\ee
where $S^{i}$ is an orientational vector satisfying $S^i S^i =1$.  This corresponds to a rotation of the $\chi$ vector in isospace. Then, upon substitution of this expression in the kinetic part of the Lagrangian (the potential depends on the modulus of $\chi$ only and therefore knows nothing of $S^i$), one obtains
\be
S_{LO} = 2\pi  \int \rho \chi(\rho)^2\; \partial^\mu S^i \partial_\mu S^i d\rho dz dt = 2\pi I_1 \int dz dt\; \partial^\mu S^i \partial_\mu S^i,
\ee
with $\mu = (t,z)$ and $I_1 = \int_0 ^\infty \rho \chi(\rho)^2 d\rho $. This is the non-linear sigma model living on the infinite vortex world sheet. In this case however the vortex is finite in length, so we must be more careful. In general we must now consider that $\chi$ is $z$ dependent. We must therefore consider
\be
\chi^i \rightarrow \chi(\rho , z) S^{i}(t,z).
\ee
This leads to the action
\be\label{act2}
S_{LO} = 2\pi  \int \rho \chi(\rho,z)^2\; \partial^\mu S^i \partial_\mu S^i d\rho dz dt = 2\pi  \int f(z)\; \partial^\mu S^i \partial_\mu S^i dz dt,
\ee
where $f(z) = \int_0^{\infty} \rho \chi(\rho,z)^2 d\rho$. There is some physics to extract here. Firstly, in the limit of $m_\chi \rightarrow \infty$ we have that $f(z)\rightarrow \Theta(z\pm z_0)$, i.e. in this limit $f(z)$ becomes a step-function localized on the monopole centers. We are exciting the string orientational degrees of freedom, effectively sending a "spin" wave along the string, but the string now has finite length so we expect there to be some tapering of the waveform. The excitations are gapless, it doesn't require a minimum amount of energy to excite this mode, but the mode must be effectively trapped due to the presence of the fixed boundaries. Let us be more precise, consider the low energy dynamics derived from the action (\ref{act2}), this is simply
\be\label{eq22}
\partial^2_t S^i - \partial^2_z S^i - \frac{\partial_z f(z)}{f(z)}\partial_z S^i =0.
\ee
When $f(z)=0$, i.e $\chi$ has no $z$ dependence and we have an infinitely long non-Abelian string then this equation is simply the free wave equation. However, the presence of a non-vanishing $f(z)$ leads to a potential-like term.  Note that $\chi$ is constant in the $z$ direction throughout the centre of the string and only varies at the end points. Therefore this tapering is only relevant at the boundaries, as it should be.  We can at first deduce the way this potential term should behave. Since $\chi$ decays exponentially in the $z$ direction beyond the monopoles $\chi \approx e^{-m_\chi z}$, with $m_\chi$ the mass of the field, then we expect
\be
\frac{\partial_z f(z)}{f(z)} \approx \pm m_\chi,
\ee
so that the potential factor should tend to a constant beyond the monopoles, the plus or minus factor depending on whether we are at positive or negative values of $z$ (clearly this is a rough estimate since there is an integration over $\rho$ involved). Between the monopoles $\chi \approx const$, so that $\partial_zf(z)=0$ and the effective potential vanishes. In fact, we can be more precise since we have an explicit numerical solution for $\chi(\rho,z)$. If we use the solution found in figure \ref{fig1} we find the $g(z) = \partial_z (\ln f(z))$ shown in figure \ref{fig3} (red). We can now use this solution to solve equation (\ref{eq22}). Firstly, the equation is separable, if we set $S^i (z,t) = q_1^i (t)q_2^i(z)$ (no summation intended) we can reduce the system to the two equations (we omit the index $i$ on the fields)
\be
\ddot{q_1}=  -\omega^2 q_1, \quad q_2''  - g(z) q_2'  = -\omega^2 q_2,
\ee
where $\omega$ is a constant. The solution to the second equation is found numerically and shown in figure \ref{fig3}. We see that the incoming perturbation, sent from $-\infty$ in the $z$-axis is severely damped outside the inter monopole spacing and excites the orientational moduli inside. The potential-like term is represented by the function $g(z)$ acting on the first derivative in $z$ of the wave functions. Let us denote this solution  as $q_{2s}(z)$. The full solution for the moduli is therefore
\be
S ^i (t,z) = q^i_{2s}(z)(c^i_1 \cos(\omega^i t)+c^i_2\sin(\omega^i t)),
\ee 
where $c_1$ and $c_2$ are constants which must obey the constraint $S^i S^i =1$. These are the gapless excitations corresponding to the orientational degrees of freedom of the NAMVC.

\subsection{The Hedgehog MVC}

In the case of the hedgehog MVC things are drastically different. The orientational moduli cease to exist. This is because the $\chi$ field does not vanish outside of the vortex core, hence the moduli are delocalized from it and it doesn't make sense to assign them as orientational moduli of the vortex.  Therefore, this solution does not represent a non-abelian monopole - vortex complex.\section{Conclusions}

This paper extends a recent proposal for a simple model of non-Abelian vortices to the NAMVC case. In particular we investigated several configurations of non-abelian vortices which end on electric sources. We found that the tension of this object is lower than its abelian counterpart, meaning the monopoles are more weakly confined. We demonstrated that a non-linear sigma model lives on the world-sheet of the soliton complex and solved for the dynamics of the orientational degrees of freedom. The $z$ dependence of the $\chi$ field effectively acts as a potential term for the low energy fields. We further found a superconducting droplet solution with a hedgehog - like ansatz for the orientational field. This configuration has no orientational degrees of freedom localised on it. Its energy is infrared divergent, meaning it must be considered in a finite box.  \newline

Unfortunately one cannot say much about real world confinement from this model. It is a gigantic assumption to suppose low energy QCD can be so neatly described by a dual weakly coupled model. However, in the context of the dual superconducting mechansim of colour confinement, something analogous to this must be happening. A low-energy remnant of the QCD degrees of freedom must condense and play the part of the $S$ field in our model. What plays the role of $\chi$, and if indeed this type of non-abelian soliton is the correct one to describe quark confinement, remains a mystery.

\subsection*{Acknowledgements}

G.T. would like to thank the University of Toronto for kind hospitality. This work has been funded by the Fondecyt grant 11160010 (G. T.) and the NSERC Discovery Grant (A. P.).

\end{document}